\documentclass[%
aip,
preprint,
amsmath,amssymb,
]{revtex4-2}

\usepackage{graphicx}
\usepackage{dcolumn}
\usepackage{bm}
\usepackage{hyperref}
\usepackage{color}
\usepackage[utf8]{inputenc}
\usepackage[T1]{fontenc}
\usepackage{mathptmx}
\usepackage{etoolbox}
\graphicspath{{./}}

\makeatletter
\def\@email#1#2{%
 \endgroup
 \patchcmd{\titleblock@produce}
  {\frontmatter@RRAPformat}
  {\frontmatter@RRAPformat{\produce@RRAP{*#1\href{mailto:#2}{#2}}}\frontmatter@RRAPformat}
  {}{}
}%
\makeatother

\newcommand{\vect}[1]{\boldsymbol{#1}}
\begin{document}

\title{Performance Evaluation of a Diamond Quantum Magnetometer for Biomagnetic Sensing: A Phantom Study}


\author{Naota Sekiguchi}
\email[]{sekiguchi.n.6ddd@m.isct.ac.jp}
\affiliation{Department of Electrical and Electronic Engineering, Institute of Science Tokyo, Meguro, Tokyo 152-8550, Japan}

\author{Yuta Kainuma}
\affiliation{Department of Electrical and Electronic Engineering, Institute of Science Tokyo, Meguro, Tokyo 152-8550, Japan}

\author{Motofumi Fushimi}
\affiliation{Department of Bioengineering, The University of Tokyo, Bunkyo, Tokyo 113-8656, Japan}

\author{Chikara Shinei}
\affiliation{Research Center for Electronic and Optical Materials, National Institute for Materials Science, Tsukuba, Ibaraki 305-0044, Japan}
\affiliation{Department of Applied Physics, University of Tsukuba, Tsukuba, Ibaraki 305-8571, Japan}

\author{Masashi Miyakawa}
\author{Takashi Taniguchi}
\affiliation{Research Center for Materials Nanoarchitectonics, National Institute for Materials Science, Tsukuba, Ibaraki 305-0044, Japan}
\author{Tokuyuki Teraji}
\affiliation{Research Center for Electronic and Optical Materials, National Institute for Materials Science, Tsukuba, Ibaraki 305-0044, Japan}

\author{Hiroshi Abe}
\author{Shinobu Onoda}
\affiliation{Takasaki Institute for Advanced Quantum Science, National Institutes for Quantum Science and Technology, Takasaki, Gunma 370-1292, Japan}

\author{Takeshi Ohshima}
\affiliation{Takasaki Institute for Advanced Quantum Science, National Institutes for Quantum Science and Technology, Takasaki, Gunma 370-1292, Japan}
\affiliation{Department of Materials Science, Tohoku University, Sendai, Miyagi 980-8579, Japan}

\author{Mutsuko Hatano}
\affiliation{Department of Electrical and Electronic Engineering, Institute of Science Tokyo, Meguro, Tokyo 152-8550, Japan}

\author{Masaki Sekino}
\affiliation{Department of Bioengineering, The University of Tokyo, Bunkyo, Tokyo 113-8656, Japan}

\author{Takayuki Iwasaki}
\affiliation{Department of Electrical and Electronic Engineering, Institute of Science Tokyo, Meguro, Tokyo 152-8550, Japan}

\date{\today}

\begin{abstract}
We employ a dry-type phantom to evaluate the performance of a diamond quantum magnetometer with a high sensitivity of about $6~\mathrm{pT/\sqrt{Hz}}$ from the viewpoint of practical measurement in biomagnetic sensing.
The dry phantom is supposed to represent an equivalent current dipole (ECD) generated by brain activity, emulating an encephalomagnetic field.
The spatial resolution of the magnetometer is evaluated to be sufficiently higher than the length of the variation in the encephalomagnetic field distribution.
The minimum detectable ECD moment is evaluated to be 0.2~nA~m by averaging about 8000 measurements for a standoff distance of 2.4~mm from the ECD.
We also discuss the feasibility of detecting an ECD in the measurement of an encephalomagnetic field in humans.
We conclude that it is feasible to detect an encephalomagnetic field from a shallow cortex area such as the primary somatosensory cortex.
\end{abstract}

\maketitle

A diamond quantum magnetometer (DQM) based on nitrogen–vacancy (NV) center ensemble in diamond is a fascinating tool for biomagnetic sensing due to its favorable characteristics \cite{Zha21, Asl23}.
DQM can be operated at room temperature with a high sensitivity currently up to $\mathrm{pT/\sqrt{Hz}}$ order \cite{Bar24, Sek24, Wan24}, which facilitates decreasing the distance to the measurement object.
The short distance leads to a better spatial resolution of the target activity \cite{Kor16} and to a significantly larger signal because the biomagnetic field typically decays faster than the $-1$ power of the distance \cite{Ham93}.
The intrinsic spatial resolution of DQM itself is determined by the optically-excited volume of NV centers and can be decreased down to the sub-mm scale \cite{Bar16, Ara22, Sek24}.
Additionally, a very wide dynamic range of DQM \cite{Cle18, Hat22, Wan23} provides the possibility of highly sensitive magnetometry in an ambient field outside a magnetic shield.

Sensitivity improvement in DQM has been actively studied \cite{Bar20} and actual application to biomagnetic sensing has been reported \cite{Bar16, Ara22, Yu24}, while few studies have reported on the evaluation of DQM from the viewpoint of biomagnetic sensing \cite{Zha23a}.
Many of those studies reported their field sensitivity achieved \cite{Bar24, Fes20, Gao23, Gra23, Sch18, Wan24, Wol15, Xie21a, Zha21a, Sek24}, but discussions about the stability \cite{Sek24} and the minimum detectable field in a biomagnetic sensing have generally been limited.
The stability is of great importance because the current sensitivity around $\sim \mathrm{pT/\sqrt{Hz}}$ in DQM generally requires a long measurement time in total for accumulating a small signal.
The minimum detectable field in this measurement depends on not only the measurement bandwidth and the number of accumulation but also characteristics of noise.
Therefore, the evaluation of the minimum detectable field is essential to infer the performance of a magnetometer in practical applications.
Furthermore, the evaluation of the intrinsic spatial resolution is important for a particular application including magnetoencephalography (MEG), where the estimation of the source generating a biomagnetic field by solving an inverse problem \cite{Ham93, Kor16} would be disturbed if the intrinsic spatial resolution is worse than the length of the variation in the field distribution to be measured.

Here, we evaluate a DQM with a high sensitivity of about  $6~\mathrm{pT/\sqrt{Hz}}$ by using a dry type of phantom for an encephalomagnetic field in small animals.
The dry phantom can emulate an encephalomagnetic field outside the brain and be considered as a representation of an equivalent current dipole (ECD) generated by brain activity.
The spatial distribution of the phantom's field was measured and agreed with the theoretical prediction, which indicates the intrinsic spatial resolution was sufficiently higher than the length of the variation in the encephalomagnetic field distribution.
The minimum detectable field and the minimum detectable ECD were investigated.
It was found that the ECD moment of about 0.5~nA~m and of 0.2~nA~m can be detected with the unity signal-to-noise ratio by averaging about 1500 and 8000 measurements, respectively, for a standoff distance of 2.4~mm.
We also discuss the feasibility of detecting an ECD in the measurement of an encephalomagnetic field in humans.
We conclude that it is feasible to detect a shallow ECD at, for example, the primary somatosensory cortex area with our DQM.

\begin{figure}[tb]
	\centering
	\includegraphics{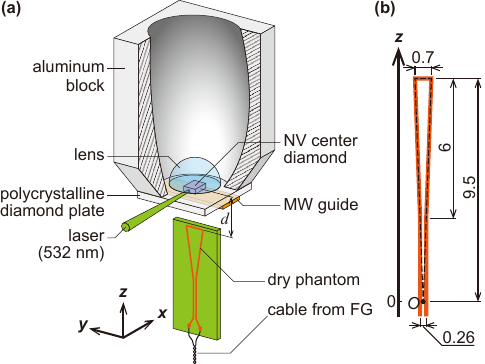}		
	\caption{\label{fig: setup}
		(a) Experimental setup (not to scale) and (b) dimensions of the designed dry phantom. 
	}
\end{figure}

A DQM used in this study is illustrated in Fig.~\ref{fig: setup}(a).
The DQM setup was almost the same as that in our previous work \cite{Sek24} and inside a magnetically-shielded room.
We used a single-crystal (111) diamond synthesized by a high-pressure and high-temperature method.
The initial concentration of substitutional nitrogen ($\mathrm{N_s^0}$) was controlled by using a titanium additive to a metal solvent as a nitrogen getter \cite{Miy22}.
NV centers were fabricated by electron beam irradiation and high-temperature annealing.
An electron spin resonance measurement yielded $[\mathrm{NV^{-}}] = 1.2~\mathrm{ppm}$ and $[\mathrm{N_s^{0}}] = 2.3~\mathrm{ppm}$.
The isotope ratio of $^{13}$C in the diamond was reduced to about 500~ppm.
The dephasing time, $T_{2}^{\ast}$, of the $\mathrm{NV^{-}}$ was estimated to be approximately $2~\mathrm{\mu s}$ \cite{Sek24}.

An ensemble of $\mathrm{NV^{-}}$ in the diamond was excited from a side face by a green laser at 532~nm with the power of 0.39~W.
The excitation laser beam was focused onto the diamond and had the spot size of about 70~µm in diameter.
The laser path length in the diamond was estimated to be 1~mm.
The intrinsic spatial resolution of this DQM was therefore estimated to be 70~µm and 1~mm along the $y$ and $x$ direction, respectively.
The laser-induced fluorescence from NV centers was collected with a hemispherical lens and an elliptically-shaped inner wall of an aluminum block and then detected by a photodiode.
The intensity noise in the fluorescence due to the excitation-laser intensity noise was reduced by a balanced detection technique \cite{Sek24}.
Heat due to the laser illumination was dissipated by attaching a polycrystalline diamond plate to the diamond.

We used a dry-type phantom that emulates an encephalomagnetic field from the brain of a small animal.
While modeling an actual magnetic field generated from a neuron is quite difficult due to the complicated currents around the neuron, an analytical formula derived by Sarvas fairly reproduces the encephalomagnetic field under the assumptions as follows:
a field source, which is supposed to be an ensemble of intracellular currents at neurons, can be approximated as a single ECD $\vect{Q}$ at $\vect{r}_0$ in the spherically symmetric conductor with its center at the origin of the coordinate.
The Sarvas' formula computes a magnetic field $\vect{B}$ at the position $\vect{r}$ of a sensor as
\begin{equation}
	\label{eq: Sarvas}
	\vect{B}(\vect{r})
	=
	\frac{\mu_0}{4\pi F^2}
	\left[
	F\vect{Q}\times\vect{r}_0
	-
	\{(\vect{Q}\times\vect{r}_0)\cdot\vect{r}\}\nabla F
	\right],
\end{equation}
where
\begin{equation}
	F = |\vect{r}-\vect{r}_0|\left(|\vect{r}-\vect{r}_0||\vect{r}| + |\vect{r}|^2 - \vect{r}_0\cdot\vect{r}\right).
\end{equation}
The dry phantom that consists of an isosceles-triangle current is known as a source generating a magnetic field obeying the Sarvas' formula \cite{Ueh07} and can emulate an encephalomagnetic field \cite{Oya15}.
Our dry phantom made on a PCB was placed below the DQM with the distance $d$ from the excited NV ensemble as shown in Fig.~\ref{fig: setup}(a).
The dimensions of the dry phantom [Fig.~\ref{fig: setup}(b)] were determined by considering the size of the head of a small animal such as a rat.
We intended to realize an isosceles-triangle coil with the base length $l$ of 0.7~mm and the leg length of 9.5~mm, while the actual legs were connected to parallel wires at 6-mm away from the base.
The ECD is supposed to be generated at the base with the moment $Q_y$ along the $y$ direction, $Q_y = i_{\mathrm{DP}}l$, where $i_{\mathrm{DP}}$ is the current flowing on the phantom.
The dry phantom was mounted on a $z$ stage to vary the distance $d$ and on a motorized $xy$ stage for the two-dimensional scan.
We applied a sinusoidal current at 33.33~Hz to this dry phantom to generate a test field.
This frequency is within the primary band of an encephalomagnetic field \cite{Har18, Uhl18}.

We performed a continuous-wave optically-detected magnetic resonance by applying a microwave (MW) current through a MW guide on the other side of the polycrystalline diamond plate.
A bias magnetic field of about 1~mT along the $z$ axis was applied to the NV center ensemble by a permanent ring magnet.
Here, the three possible resonances, which are associated with the hyperfine manifold, between the electron ground states $|0\rangle \leftrightarrow |1\rangle$ were simultaneously driven by using a three-tone MW field \cite{Bar16}.
The frequency of the three-tone field was modulated to employ lock-in detection.
During the measurement of a magnetic field, we stabilized the MW frequency to the resonance frequency by monitoring the lock-in signal  $S_{\mathrm{LI}}$ and applying a slow (2~Hz) PID servo to an MW generator.
A magnetic field $B_{\mathrm{m}}$ faster than the PID servo was given as $B_{\mathrm{m}}=S_{\mathrm{LI}} / (dS_{\mathrm{LI}}/df \times \gamma_e)$, where $dS_{\mathrm{LI}}/df$ is the zero-crossing slope around the resonance and $\gamma_e=28~\mathrm{GHz/T}$ is the gyromagnetic ratio.
The minimum distance between the excited NV ensemble and a measurement object was estimated to be 0.8~mm, limited by the thicknesses of the NV center diamond, the heat spreading plate, and the MW guide.

\begin{figure}[t]
	\centering
	\includegraphics{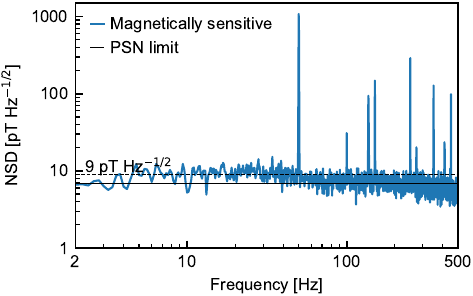}		
	\caption{\label{fig: NSD}
		Single-sided noise spectral density. 
		The dashed line indicates the noise density floor of $9~\mathrm{pT~Hz^{-1/2}}$ at 33.33~Hz.
	}
\end{figure}

The single-sided noise spectral density in the DQM was measured with the cut-off frequency of 500-Hz at the lock-in detection as shown in Fig.~\ref{fig: NSD}.
The corresponding single-sided noise spectral density determined by photon shot-noise was calculated to be $6.9~\mathrm{pT~Hz^{-1/2}}$ (black solid line).
The measured noise density was found to be almost limited by the photon shot noise.
We also found that no noise peaks other than the power-line noises at 50~Hz and its harmonics were observed at the encephalomagnetic-field frequency band $< 100~\mathrm{Hz}$ \cite{Har18, Uhl18}.
The single-sided noise density floor around the test-field frequency of 33.33~Hz was estimated to be $9~\mathrm{pT~Hz^{-1/2}}$ (dashed line), which corresponds to the field sensitivity of $9~\mathrm{pT~Hz^{-1/2}} / \sqrt{2}=6~\mathrm{pT~Hz^{-1/2}}$.

\begin{figure*}
	\centering
	\includegraphics{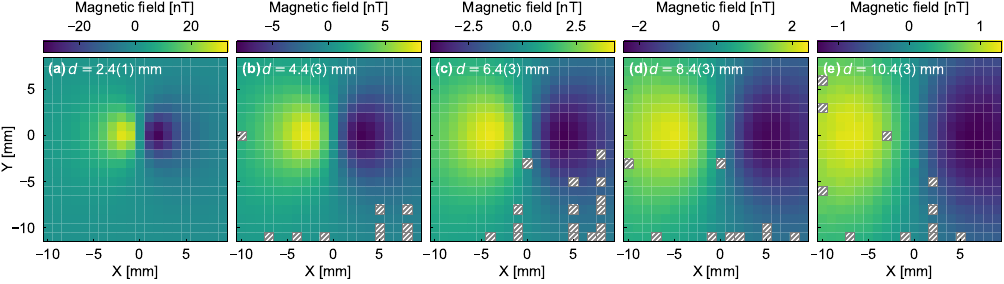}		
	\caption{\label{fig: mapping}
		Mapped magnetic field along the $z$ axis with the different distance $d$.
		The measurement at the hatched regions failed due to a large noise from the motorized stage.
		The center of the phantom's base was aligned to the diamond center at $(x, y)=(0, 0)$ by the eye.
	}
\end{figure*}
\begin{figure}
	\centering
	\includegraphics{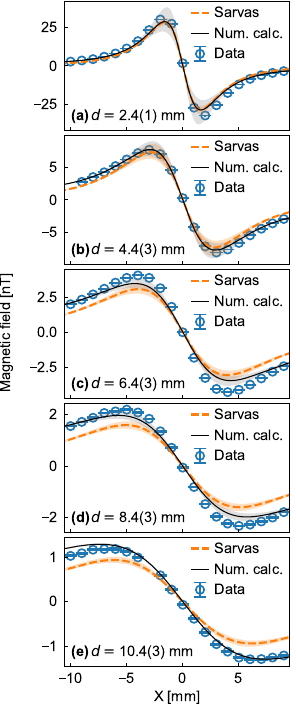}
	\caption{\label{fig: profile}
		Magnetic field profile along the $x$ axis at $y=0$ for the different distance $d$.
		The measured field is shown by the blue circles.
		The black solid and orange dashed lines represent the numerically calculated profiles respectively by the Biot-Savart law and by the Sarvas' formula.
		The colored bands about the lines indicate the uncertainty caused by the uncertainty in $d$.
	}
\end{figure}

We mapped the $z$ component of the test field generated by the dry phantom by scanning the $x$ and $y$ positions of the phantom at five different distances as shown in Fig.~\ref{fig: mapping}.
The heatmaps show the amplitude of the sinusoidal test field obtained by least square method for the current peak amplitude of 7.19~mA.
At some positions (the hatched points), the measurement failed because the motorized stage generated a magnetic field during the scanning that was larger than the applicable range of the PID servo for the MW frequency.
We observed two peaks with opposite polarity and the sharper distribution for the closer distance, as expected from the ECD.
The distance between the two peaks approximately followed the distance $d$.
The peak amplitude dependence on $d$ shows the decay faster than $d^{-1}$.

The field profiles along $x$ axis at $y=0$ were extracted to compare the measured field with the numerically calculated field by the Biot-Savart law and the Sarvas' formula [Eq.~\eqref{eq: Sarvas}].
Figure~\ref{fig: profile}(a)-(e) corresponds to the profile for Fig.~\ref{fig: mapping}(a)-(e), respectively.
We found that the numerically calculated field (black solid line) by the Biot-Savart law with the current of 7.19~mA along the designed phantom's path agreed with the measured data (blue points) at all distances.
This suggests that the intrinsic spatial resolution of about 1~mm along the $x$ direction was sufficient to resolve the peaks of the phantom's field at $d=2.4\pm0.1~\mathrm{mm}$, which is comparable to or shorter than the typical standoff distance from the cortex of a small animal such as a rat \cite{Pax98}.
The discrepancy between the numerically calculated field and the field given by the Sarvas's formula (orange dashed line) at larger distances indicates imperfection in our dry phantom.
It was found by numerical calculation that the discrepancy was attributed to two factors:
the absence of the V-shaped wire compared to the ideal isosceles triangle; 
and the presence of the slant wires from the parallel wire to pads for connecting a cable [see Fig.~\ref{fig: setup}(a)].

We investigated the minimum detectable magnetic field $B_{\mathrm{md}}$ in a typical measurement of biomagnetism, where one repeatedly acquires time traces of a biomagnetic field signal and averages it.
In this measurement, the phantom was located at $(x, y) = (2.0, 0.0)~\mathrm{mm}$ and $d=2.4~\mathrm{mm}$.
The time trace for a period of 570~ms was acquired 8000 times.
The total measurement time was 76 minutes.
The cut-off frequency of the low-pass filter at the lock-in amplifier was set to 100~Hz.
The sinusoidal test current was sent at the time $t = 200~\mathrm{ms}$ for 8 periods to the phantom.
The peak amplitude of the current was $i_{\mathrm{DP}} = 0.69~\mathrm{\mu A}$.
The corresponding moment of an ECD was roughly estimated to be $Q_y=0.5~\mathrm{nA~m}$.
It is supposed that small animals like rats generate such a small current dipole in the brain cortex by stimulation to, for example, auditory \cite{Kim17}.
A type of spontaneous brain activity including epilepsy should produce a stronger ECD \cite{Che92, Ale14}, but it is difficult to average the field generated.

\begin{figure}
	\centering
	\includegraphics{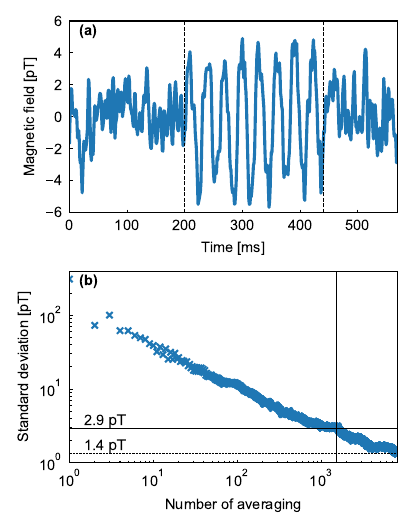}
	\caption{\label{fig: trace}
		Time domain measurement at $(x, y)=(2.0, 0.0)~\mathrm{mm}$.
		(a) Time trace averaged over 8000 measurements.
		The test field was generated at $t=200~\mathrm{ms}$ for 8 periods (between the dashed lines).
		(b) The standard deviation at $t<200~\mathrm{ms}$ as a function of the number of averaging.
	}
\end{figure}

The time trace averaged over the 8000 acquisitions is shown in Fig.~\ref{fig: trace}(a).
The phantom's test field was clearly observed and measured to have the root-mean-square amplitude of $2.9~\mathrm{pT}$, while the standard deviation ($B_{\mathrm{md}}$) at $t<200~\mathrm{ms}$ was measured to be 1.4~pT.
Furthermore, the minimum detectable ECD moment for the 8000 times averaging and $d=2.4~\mathrm{mm}$ can be estimated to be $0.5~\mathrm{nA~m} / 2.1 \simeq 0.2~\mathrm{nA~m}$.

We analyzed the decrease in the standard deviation in an averaged trace at $t<200~\mathrm{ms}$ as the number of averaging was increased, as shown in Fig.~\ref{fig: trace}(b).
It was found that the standard deviation scales as the number of averaging to the power of $-1/2$, which was enabled by the good noise performance.
Since the standard deviation reached 2.9~pT at around 1500 times averaging, we consider that our DQM can detect an ECD with $Q_y\sim 0.5~\mathrm{nA~m}$ by 15-minute measurement.

\begin{figure}
	\centering
	\includegraphics{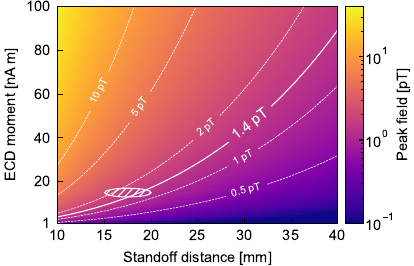}
	\caption{\label{fig: det_ecd}
		Simulated peak field as a function of the standoff distance and the ECD moment.
		The white lines are the contour lines of the peak field.
		The hatched circle represents the previously reported ECD moment and standoff distance \cite{Kor13}.
	}
\end{figure}

Furthermore, we investigated the detectable ECD moment based on the Sarvas' formula for the case of encephalomagnetic-field measurement of a human.
Here, it was assumed for simplicity that the ECD was located on the $z$ axis at $z_0$ and oriented along the tangential direction ($x$-$y$ plane).
Considering the size of a human brain, the distance between the center of the head-model sphere and the DQM was fixed to 100~mm in this numerical calculation.
The $z$ component of the field was numerically calculated by Eq.~\eqref{eq: Sarvas} and mapped by scanning the DQM position in the $x$-$y$ plane.
We explored the field maximum $B_{z, \mathrm{max}}$ in the calculated field distribution for a given ECD moment strength.
Figure~\ref{fig: det_ecd} shows $B_{z, \mathrm{max}}$ as a function of the standoff distance and the ECD moment.
The standoff distance may be limited by the depth of the ECD from the head surface, since the measurement distance of the DQM can be decreased to about 1~mm and negligible compared to the depth.
The minimum detectable field $B_{\mathrm{md}}=1.4~\mathrm{pT}$ for the case of 8000 averaging is indicated by the white solid contour line.
The other contours in dashed line are guides for the eye.
It is supposed that a typical moment of an ECD in MEG measurement would be on the order of 10~nA~m \cite{Ham93}.
The standoff distance largely varies on the activated region where the ECD is generated.
For example, for somatosensory stimulation, an ECD with the moment of about 15~nA~m at the depth of 15-20~mm (hatched circle) has been reported \cite{Kor13}.
This simulation showed that the peak field of the encephalomagnetic field that is generated by the previously reported ECD is stronger than the minimum detectable field of 1.4 pT.
Therefore, it is feasible for our DQM to detect an encephalomagnetic field from such a shallow ECDs in humans.

We performed the measurement of a magnetic field generated by a dry-type phantom that emulated an encephalomagnetic field in a small animal to evaluate a highly sensitive DQM from the viewpoint of biomagnetic sensing.
The single-sided noise spectral density of the DQM showed the very low noise floor of $9~\mathrm{pT~Hz^{-1/2}}$, which corresponds to the sensitivity of about $6~\mathrm{pT~Hz^{-1/2}}$.
The spatial distribution of the phantom's field was measured by scanning the phantom relative to the DQM.
The intrinsic spatial resolution of about 1~mm along the $x$ direction of the DQM enabled us to observe the clear peaks of the phantom's field without smearing.
For the case of time domain measurement, the minimum detectable field was found to be 1.4~pT with about 100-Hz bandwidth by averaging signal 8000 times.
It was found that the ECD moment of about about 0.2~nA~m can be detected by averaging about 8000 measurements at the standoff distance of 2.4~mm.
We also considered that it is feasible to detect an encephalomagnetic field from a shallow region of a human brain like the primary somatosensory cortex area by using our DQM.
Although the estimations of the position and the moment of an ECD were not performed due to the discrepancy of our dry phantom from an ideal one, it is demonstrated that the evaluation of a DQM using a phantom is beneficial from the viewpoint of biomagnetic sensing.

This work was supported by the MEXT Quantum Leap Flagship Program (MEXT Q-LEAP) Grant No. JPMXS0118067395 and JPMXS0118068379.

\bibliography{DP_DQM}

\end{document}